\documentclass[12pt]{iopart}
\usepackage{mathrsfs}
\usepackage{graphicx}

\usepackage{iopams}

\def\be{\begin{equation}}
\def\ee{\end{equation}}
\def\bea{\begin{eqnarray}}
\def\eea{\end{eqnarray}}

\begin{document}

\title[]{On black hole spectroscopy via quantum tunneling}
\author{Qing-Quan Jiang$^1$\footnote{Corresponding author}}
\address{$^1$College of Physics and Electronic Information, China
 West Normal University, Nanchong, Sichuan 637002, People's Republic of China}
 \ead{qqjiangphys@yeah.net}

\begin{abstract}
In recent work [\emph{Quantum tunneling and black hole spectroscopy, Phys. Lett.} B686 (2010) 279], it has been shown, in the tunneling mechanism, the area spacing parameter of a black hole horizon is given by $\gamma=4$. In this paper, by carefully analyzing the tunneling process of the black hole radiation, we interestingly find that the most qualified candidate for a universal area gap in the tunneling mechanism is $\gamma=8\pi$.
First, we develop the Banerjee's treatment and the Kunstatter's conjecture to revisit the black hole spectroscopy via quantum tunneling, and find for a real tunneling process, the area spacing parameter is given by the possible value $\gamma\geq 4$. That is, the previous model-dependent area spacing parameters, i.e. $\gamma=8\pi, 4\ln 3, 4$, are all possible in the tunneling mechanism. Finally, some discussions are followed to find, in the tunneling mechanism, $\gamma=8\pi$ is the most qualified candidate for a universal area spacing parameter.

\end{abstract}

\pacs{04.70.Dy, 03.65.Sq}
\maketitle

\section{Introduction}

During the last forty years evidence has been mounting that the horizon area of a black hole has a discrete spectrum.
Moreover, it is widely believed that the area eigenvalues are uniformly spaced, i.e.
\begin{equation}
A_n=\gamma l_p^2 n, \label{eq1}
\end{equation}
where $l_p$ is the Planck length, and $\gamma$ is regarded as a numerical coefficient of the order of unity.
However, there is no general agreement on the spacing of the levels, that is, the value of $\gamma$ has been somewhat controversial. On this point, Bekenstein was the first one to find, when a neutral particle is absorbed by a black hole, the horizon area is universally increased by $\gamma=8\pi$ \cite{Bekenstein1972,Bekenstein1973}. Since then, many treatments have independently reproduced
this universal value. These examples include the Maggiore's reinterpretation \cite{r6} of the renowned Hod conjecture \cite{r3}, a quantization procedure proposed by Ropotenko \cite{rrrr4}, a refinement thereof \cite{Medved1},  a method reducing the black hole phase space to a pair of observable \cite{nrrr4}, identifying the exponent of the gravitational action as a quantum amplitude \cite{Padmanabhan1}, applying non-commutative quantum theory \cite{Romero1,Dolan1}, and a recent attempt to use the adiabatic invariance \cite{r7} \footnote{most applications of these treatments see in \cite{NEW1,NNEW1,NNEW3,NNEW4,Fernando79,Chen69,
Kwon28,Daghigh26,Kwon27,Kwon27165011,Wei2010,Lopez-Ortega,MyungarXiv:1003.3519}}, etc. Of course, various schools of thought have questioned this value of the area spacing parameter. For example, Hod \cite{r3}, Kunstatter \cite{r4} and others \cite{Barvinsky2001,Bekenstein2002,Bekenstein200266,NEW2,NNEW2} suggested, in the context of quasinormal modes, the area spacing parameter of a black hole horizon should be given by $\gamma=4\ln 3$. And a recent observation by Banerjee, Majhi and Vagenas \cite{rrr2} has shown, in the tunneling mechanism, the area spacing parameter of a black hole horizon is fixed by $\gamma=4 $ \footnote{Following this idea, some related work \cite{rrrr2,rrrr3,rrrr5,rrrr6} have appeared to obtain the same area spacing.}. In a word, there is no general agreement on the area spacing parameter, and $\gamma=8\pi, 4\ln 3, 4$ are all possible. In \cite{r8,rrrr7}, Medved commented on this inconsistency, and declared $\gamma=8\pi$ was still, by far, the most qualified candidate for a universal area spacing.

On the other hand, some recent work \cite{rrr2,rrrr2,rrrr3,rrrr5,rrrr6} has shown, in the tunneling mechanism, the area spacing parameter of a black hole horizon is fixed by $\gamma=4$.
This spacing level is encouraging since it is not only in full agreement with the Hod's result by considering the capture of a quantum (finite size) charged particle by a black hole \cite{Hod2}, but more importantly, is smaller than that given by Bekenstein \cite{Bekenstein1972} for neutral particles as well as the one computed in the context of black hole quasinormal modes \cite{r6}. However, it is unconvincing since the tunneling mechanism used in these work cannot truly describe the emission process of a black hole. In the tunneling mechanism, when a particle tunnels out or in, because of the negative heat capacity, an evaporating black hole (a Schwarzschild black hole or a Kerr-Newman black hole) is (when in isolation) a highly unstable system, which means the thermal equilibrium between the black hole and the outside is unstable. That is, there would be a difference in temperature between them. Under this notion, we conclude the emission process of a black hole should be irreversible. Unfortunately, in recent work \cite{rrr2,rrrr2,rrrr3,rrrr5,rrrr6}, the tunneling process has all been treated as a reversible process, where the black hole and the outside approach a thermal equilibrium when a particle tunnels out or in. Hence, the area spacing parameter $\gamma=4$ obtained in the tunneling mechanism, is only suitable for a reversible process. In this paper,
we revisit the black hole spectroscopy in the tunneling mechanism, and attempt to find the most qualified area spacing parameter for a real tunneling process.

This paper is organized as follows: In Sec. \ref{sec1}, to prepare for the following research, we first calculate the expectation value of the emitted particle in the tunneling picture. By carefully analyzing the emission process of a black hole, in Sec. \ref{sec2}, we use the Banerjee's treatment to revisit the black hole spectroscopy in the tunneling mechanism. In Sec. \ref{sec3}, associating the vibrational frequency of a black hole with its transition frequency in large number $n$, we refine the Kunstatter's conjecture to confirm our observation in Sec. \ref{sec2}. Sec. \ref{sec4} ends up with a discussion on the most qualified area spacing from the extreme conditions ($n\gg 1$ and $n=1$), and a conclusion.

\section{Quantum tunneling mechanism}     \label{sec1}

 In this section we briefly review the modified tunneling method as developed by Banerjee and Majhi \cite{Banerjee1,Banerjee2}, to produce the expectation value of the emitted particle $\langle\omega\rangle$. For simplicity, we consider a black hole characterised by a spherically symmetric, static space-time and asymptotically flat metric of the form,
\be
ds^2=f(r)dt^2-\frac{1}{f(r)}dr^2-r^2(d\theta^2+\sin^2\theta d\phi^2), \label{eq11}
\ee
where the black hole horizon $r_h$ is determined by $f(r_h)=0$. Now, we focus our attention on studying quantum tunneling from the black hole horizon. It should be noted that, since the emission behavior across the horizon is radially tunneling, it is enough to consider the $r-t$ sector of the spacetimes. Considering the massless scalar particle governed by the Klein-Gordon equation
\be
-\frac{\hbar^2}{\sqrt{-g}}\partial_\mu[g^{\mu\nu}\sqrt{-g}\partial_\nu]\Phi(r,t)=0,
\ee
and choosing the standard (WKB) ansatz as
\be
\Phi(r,t)=\exp\big[{-\frac{i}{\hbar}S(r,t)}\big],
\ee
where
\be
S(r,t)=S_0(r,t)+\sum_{i=1}^\infty\hbar^iS_i(r,t),
\ee
in the semiclassical limit (i.e. $\hbar\rightarrow 0$), we obtain
\begin{equation}
\frac{\partial S_0(r,t)}{\partial t}=\pm f(r)\frac{\partial S_0(r,t)}{\partial r}.
\end{equation}
For the metric (\ref{eq11}), it has a timelike Killing vector, so the semiclassical action can be written as $S_0(r,t)=\omega t+S_0(r)$, where $\omega$ is the conserved quantity corresponding to the time translational Killing vector field, which is identified
as the effective energy experienced by the particle at asymptotic infinity. In this case, one can easily find the solution for $S_0(r)$. Finally, the semiclassical action is determined by
\begin{equation}
S_0(r,t)=\omega(t\pm r^\ast), ~~~~r^\ast=\int \frac{dr}{f(r)}.\label{eq13}
\end{equation}
Hence, the solution for the scalar field $\Phi$ can be written as
\begin{equation}
\Phi(r,t)=\exp\Big[{-\frac{i}{\hbar}\omega(t\pm r^\ast)}\Big]. \label{eq14}
\end{equation}
Now, we define the right moving mode, when acting as the eigenstate of the radial momentum operator $\widehat{p}(r)$, whose eigenvalue is positive, while the left moving mode corresponds to a negative eigenvalue. Thus the right and left modes inside and outside the black hole horizon are read off
\begin{eqnarray}
&&\Phi_{\textrm{in}}^{R}=\exp\Big(-\frac{i}{\hbar}\omega u_{\textrm{in}}\Big); ~~~ \Phi_{\textrm{out}}^{R}=\exp\Big(-\frac{i}{\hbar}\omega u_{\textrm{out}}\Big);\nonumber\\
&&\Phi_{\textrm{in}}^{L}=\exp\Big(-\frac{i}{\hbar}\omega v_{\textrm{in}}\Big);~~~\Phi_{\textrm{out}}^{L}=\exp\Big(-\frac{i}{\hbar}\omega v_{\textrm{out}}\Big). \label{eq15}
\end{eqnarray}
Here, $v=t+r^{*}$ and $u=t-r^*$ denote the advanced and retarded coordinates, respectively. To connect the right and left moving modes defined inside and outside the horizon, we should first find a coordinate system in which the metric (\ref{eq11}) is defined both inside and outside the horizon. In \cite{Banerjee1,Banerjee2}, through defining the Kruskal time ($T$) and space ($X$) inside and outside the horizon as
\begin{eqnarray}
&&T_{\textrm{in}}=\exp(\kappa r^\ast_{\textrm{in}})\cosh(\kappa t_{\textrm{in}}); ~~~ X_{\textrm{in}}=\exp(\kappa r^\ast_{\textrm{in}})\sinh(\kappa t_{\textrm{in}});\nonumber\\
&&T_{\textrm{out}}=\exp(\kappa r^\ast_{\textrm{out}})\sinh(\kappa t_{\textrm{out}});~~~X_{\textrm{out}}=\exp(\kappa r^\ast_{\textrm{out}})\cosh(\kappa t_{\textrm{out}}), \label{neq15}
\end{eqnarray}
one has provided a variable metric on both sides of the horizon. Here, $\kappa=\frac{1}{2}\partial_rf(r)|_{r=r_h}$ is the surface gravity of the black hole. In this case, the time ($t$) and space ($r$) inside and outside the horizon are related by
\be
t_{\textrm{in}}\rightarrow t_{\textrm{out}}-\frac{i\pi}{2\kappa},~~~~r^*_{\textrm{in}}\rightarrow r^*_{\textrm{out}}+\frac{i\pi}{2\kappa}. \label{eq16}
\ee
With this mapping, $T_{\textrm{in}}\rightarrow T_{\textrm{out}}$ and $X_{\textrm{in}}\rightarrow X_{\textrm{out}}$. Now, following the definition (\ref{eq15}), the right and left modes inside and outside the horizon are connected by
\be
\Phi_{\textrm{in}}^{L}\rightarrow \Phi_{\textrm{out}}^{L}, ~~~~~~\Phi_{\textrm{in}}^{R}\rightarrow \exp({-\frac{\pi \omega}{\hbar\kappa}})\Phi_{\textrm{out}}^{R}. \label{eq17}
\ee
In the tunneling picture, when pair production occurs inside the horizon, the left mode is trapped inside the horizon, while the right mode can tunnel across the horizon to be observed at the asymptotic infinity. Thus, the average value of $\omega$ is given by
\begin{equation}
\langle\omega\rangle=\frac{\int_0^\infty(\Phi_{\textrm{in}}^{R})^*\omega\Phi_{\textrm{in}}^{R}d\omega}
{\int_0^\infty(\Phi_{\textrm{in}}^{R})^*\Phi_{\textrm{in}}^{R}d\omega}.\label{eq18}
\end{equation}
Here, the average value of the energy $\omega$ is related to the right mode inside the horizon. Actually, the observer is lived outside the horizon, so it is necessary to recast the ``in"  quantity into its ``out" correspondence, which could easily be done using (\ref{eq17}). Thus, we have
\begin{equation}
\langle\omega\rangle=\frac{\int_0^\infty \exp{\big(-\frac{\omega}{T_h}\big)}(\Phi_{\textrm{out}}^{R})^*\omega\Phi_{\textrm{out}}^{R}d\omega}
{\int_0^\infty\exp{\big(-\frac{\omega}{T_h}\big)}(\Phi_{\textrm{out}}^{R})^*\Phi_{\textrm{out}}^{R}d\omega}
=T_h,\label{eq19}
\end{equation}
where, $T_h=\frac{\hbar \kappa}{2\pi}$, is the Hawking temperature of the black hole. In particular, for the Schwarzschild black hole, $T_h=\frac{\hbar}{8\pi M}$. This is the average energy of the emitted particle. Basing on it, we can use the Banerjee's treatment and the Kunstatter's conjecture to revisit the black hole spectroscopy in the tunneling mechanism.

\section{The Banerjee's treatment and black hole spectroscopy } \label{sec2}

In recent work \cite{rrr2}, Banerjee et al found, in the tunneling mechanism, the area spacing parameter of a black hole horizon is given by $\gamma=4$. Later, some efforts on this direction are followed to fix the same area spacing \cite{rrrr2,rrrr3,rrrr5,rrrr6}. However, this spacing level is unconvincing since the tunneling mechanism used in these work cannot truly describe the emission process of a black hole. In this section, by carefully analyzing the emission process of a black hole, we use the treatment by Banerjee et al to revisit the black hole spectroscopy. In their work, the emitted particle's energy has been treated as the lack of information in energy of the black
hole due to the particle's emission. Also,
since in information theory the entropy is lack of information, then the first law of black hole
thermodynamics can be exploited to connect these quantities. On the other hand, we note that the first law of black hole
thermodynamics, $\frac{d\omega}{T}=dS$, is an incorporation of the energy conservation law, $d\omega=dQ$, and the second law of thermodynamics, $dS = \frac{dQ}{T}$. The equation of energy conservation is suitable for any process (reversible or irreversible process). But the
equation $dS = \frac{dQ}{T}$ is true only for a reversible process. For an irreversible process, we have $dS > \frac{dQ}{T}$. That is, the tunneling mechanism involved in \cite{rrr2,rrrr2,rrrr3,rrrr5,rrrr6} has treated the emission process as a reversible process. In their treatment, the black hole
and the outside approach an thermal equilibrium during the tunneling process. But, in fact, because of the negative heat capacity, an evaporating black hole (a Schwarzschild black hole or a Kerr-Newman black hole) is (when in isolation) a highly unstable system. When a particle tunnels out or in, the thermal equilibrium between the black hole and the outside is unstable. There will be a difference in temperature, that is, a real tunneling process is irreversible. Under this notion, for any real tunneling process, the first law of black hole thermodynamics should be given by
\be
dS\geq \frac{d\omega}{T_h}, \label{eq20}
\ee
where $\omega$ is the mass or rest energy of the black hole. When a particle with energy $\langle\omega\rangle$ tunnels out from the black hole horizon, we have
\be
d\omega=\langle\omega\rangle.\label{eq21}
\ee
It should be noted that the authors of \cite{rrr2} take the change in black hole energy to be $\Delta \omega=\sqrt{\langle\omega^2\rangle-\langle\omega\rangle^2}$, which does not seem quite right. However, since $\Delta \omega=\langle\omega\rangle=T_h$, their choice amounts to the same thing. Now, substituting (\ref{eq21}) into (\ref{eq20}), we have
\be
dS\geq \frac{\langle\omega\rangle}{T_h}. \label{eq22}
\ee
In Einstein theory, the entropy of a black hole is given by the Bekenstein-Hawking
formula, $S=\frac{A}{4l_p^2}$. So, the area pacing parameter is given by
 \be
\gamma\geq 4. \label{eq23}
\ee
Obviously, in \cite{rrr2}, or later developed in \cite{rrrr2,rrrr3,rrrr5,rrrr6}, the authors only present a lower bound of the area spacing. In fact, when considering a real emission process of an evaporating black hole, we find the area spacing parameter is given by the possible value $\gamma\geq 4$. That is, the three well-known types of the area spacing parameters, i.e. $\gamma=8\pi, 4\ln 3, 4$, are all possible in the tunneling mechanism. In Sec. \ref{sec4}, by further analyzing the extreme conditions ($n\gg 1$ and $n=1$), we will observe, $\gamma=8\pi$, remains, by far, the most qualified candidate for a universal area spacing parameter.

\section{The Kunstatter's conjecture and black hole spectroscopy }\label{sec3}

In \cite{r4}, Kunstatter combined the proposal by Bekenstein for the adiabaticity of the black hole area, with the conjecture by Hod for the relation between the frequencies of the quasinormal modes and the vibrational frequencies of the black hole, to obtain the black hole spectroscopy in the context of quasinormal modes. In this section, we attempt to develop the Kunstatter's conjecture to the case of the tunneling mechanism. That is, we attempt to obtain the black hole spectroscopy by combining the adiabaticity of the black hole area with the tunneling mechanism. As stated in \cite{r4}, for a system with energy $\omega$ and vibrational frequency $f(\omega)$, it is a straightforward exercise to show that the quantity
$I=\int \frac{d\omega}{f(\omega)}$
is an adiabatic invariant. In particular, for a black hole, if $f_{\textrm{black}}(\omega)$ is the vibrational frequency of the black hole, the adiabatic invariant is read off
\be
I_{\textrm{adia}}=\int \frac{d\omega}{f_{\textrm{black}}(\omega)}. \label{eq24}
\ee
Obviously, to quantize a black hole horizon by using the adiabatic invariant (\ref{eq24}), a key point is to find the vibrational frequency $f_{\textrm{black}}(\omega)$. In the tunneling framework, the energy of the emitted particle is given by (\ref{eq19}). This means the characteristic frequency of the
outgoing particle is
\be
f_{\textrm{particle}}=\frac{\langle\omega\rangle}{\hbar}=\frac{T_h}{\hbar}. \label{eq25}
\ee
We also note that $f_{\textrm{particle}}$ is the characteristic frequency of the
outgoing particle, which corresponds to the transition frequency of a quantized black hole from a large $n$ state to another. On the other hand, in large limit ($n\gg 1$), in the spirit of the Bohr's correspondence principle, this transition frequency corresponds to the vibrational frequency of the black hole, that is
\be
f_{\textrm{black}}=f_{\textrm{particle}}.\label{eq26}
\ee
Substituting (\ref{eq25}) and (\ref{eq26}) into (\ref{eq24}), we have
\be
I_{\textrm{adia}}=\hbar\int \frac{d\omega}{T_h}. \label{eq27}
\ee
As described in Sec. \ref{sec2}, when a particle tunnels out from the black hole horizon, the entropy and the energy are connected by (\ref{eq20}) for a  real emission process. Substituting it into (\ref{eq27}), we have
\be
I_{\textrm{adia}}\leq \hbar S. \label{eq28}
\ee
Also, in the semiclassical (large $n$) limit, the adiabatic invariant $I_{\textrm{adia}}$ has an equally spaced spectrum, i.e.
\be
I_{\textrm{adia}}=n\hbar. \label{eq29}
\ee
Hence, combining (\ref{eq28}) and (\ref{eq29}), we can immediately infer that the entropy spectrum is given by
\be
S\geq n. \label{eq30}
\ee
In Einstein theory, the entropy of a black hole is given by the Bekenstein-Hawking
formula, $S=\frac{A}{4l_p^2}$. So, the area of the black hole horizon is also quantized with the area spacing parameter given by
\be
\gamma\geq 4. \label{eq31}
\ee
In the tunneling mechanism, by analyzing the black hole adiabaticity, we also find the area of the black hole horizon is quantized, with the possible area spacing parameter given by $\gamma\geq 4$, as stated in Sec. \ref{sec2}. Obviously, the previous inconsistency between the model-dependent area spacing parameters, i.e. $\gamma=8\pi, 4\ln 3, 4$, is reconciled in the tunneling mechanism. That is, the three well-known types of the area spacing parameters are all possible in the tunneling mechanism. However, for a quantum gravity theory, only one of them is expected to be in effect. In the following section, we attempt to obtain it from the extreme conditions ($n\gg 1$ and $n=1$).

\section{Conclusions and Discussions}\label{sec4}
In this paper, by carefully analyzing the emission process of a black hole, we develop the Banerjee's treatment and the Kunstatter's conjecture to revisit the black hole spectroscopy via quantum tunneling. The result shows, in the tunneling mechanism, the area of the black hole horizon is quantized evenly, with the possible spacing parameters given by $\gamma\geq 4$. In contrast, the area spacing parameter $\gamma= 4$  obtained in previous work \cite{rrr2,rrrr2,rrrr3,rrrr5,rrrr6} is only suitable for a reversible process. Interestingly, the previous inconsistency between the model-dependent area spacing parameters, i.e. $\gamma=8\pi, 4\ln 3, 4$, is reconciled in the tunneling mechanism. That is, the three well-known types of the area spacing parameters are all possible in the tunneling picture. In fact, this observation is not a bit surprising. In previous work by Hod \cite{Hod2}, a similar area spacing
was reproduced by analyzing the capture of a quantum (finite size) charged particle by a black hole.
There, the Schwinger mechanism was supplemented for a charged particle to semiclassically quantize a black hole horizon. On the other hand, we have noted that the tunneling mechanism behaves very similar with the Schwinger mechanism \cite{Srinivasan2,Kim2}. Hence, it is not surprising that, in the tunneling mechanism, a black hole is endowed with the area spacing parameters given by the possible values $\gamma\geq 4$.
However, for a quantum gravity theory, only one of them is expected to be in effect. Next, we attempt to obtain it from the extreme conditions ($n\gg 1$ and $n=1$).

Now, from the extreme conditions, i.e. $n\gg 1$ and $n=1$, we comment on the possible area spacing parameters, i.e. $\gamma=8\pi, 4\ln 3, 4$. Before that, we first find the mass spectrum of a quantized black hole. For simplicity, we consider a Schwarzschild black hole without loss of generality. Given a Schwarzschild black hole with an equally spacing area spectrum given by (\ref{eq1}), then the mass spectrum reads off
\be
M^2=\frac{\gamma}{16\pi}l_p^2 n. \label{eq32}
\ee

\textbf{i)} In the classical condition, i.e. $n\gg 1$, the energy spacing between states for which $n$ differs by one is
\be
\Delta M=\frac{\gamma}{32\pi M}l_p^2.
\ee
The corresponding transition frequency is then given by
\be
\omega_t=\frac{\Delta M}{\hbar}=\frac{\gamma}{32\pi M}.
\ee
On the other hand, it is well-known that the gravity system is a periodic system, and the corresponding period is the geometric origin of Hawking thermal radiation \cite{Gibbons2}. For a Schwarzschild black hole, the period is given by
$T=\frac{\hbar}{T_h}=8\pi M$, so the corresponding classical frequency is
\be
\omega_c=\frac{2\pi}{T}=\frac{1}{4M}.
\ee
In the spirit of the Bohr's correspondence principle, that is ``transition frequencies at large quantum number corresponds to classical oscillation frequencies", we have
\be
\omega_t=\omega_c,
\ee
which yields
\be
\gamma=8\pi. \label{eq33}
\ee
Obviously, if the possible area spacing parameter in the tunneling mechanism is fixed by $\gamma=8\pi$, it satisfies the Bohr's correspondence principle.

\textbf{ii)} For all allowed states, if $M$ is correctly given by (\ref{eq32}), the surface area of the hole exceeds $\frac{\gamma^2}{16\pi}\left(\frac{\hbar}{M}\right)^2$, except for the
ground state, i.e. $n=1$, for which the surface area is given by $A=\frac{\gamma^2}{16\pi}\left(\frac{\hbar}{M}\right)^2$. On the other hand, for any quantum object, it cannot be localized to better than its Comptom wavelength. That is, for a black hole with mass $M$, the surface area must exceed $4\pi \left(\frac{\hbar}{M}\right)^2$. This also emphasizes
\be
\gamma=8\pi. \label{eq34}
\ee
Evidently, if the area spacing parameter in the tunneling mechanism is described by $\gamma=8\pi$, we can find a satisfying result: the Schwarzschild black hole is always larger than its Compton wavelength, except in its ground
state when it is of just the same size. In this case, the ground state may be regarded also as
an elementary particle. To conclude, in the tunneling mechanism, $\gamma=8\pi$ is the most qualified area spacing for a universal area gap.

\section*{Acknowledgments}
This work is supported by the National Natural Science Foundation of China with Grant No.
11005086, and by the Sichuan Youth Science and Technology
Foundation with Grant No. 2011JQ0019, and by a starting fund of China West Normal University with Grant No. 10B016.

\end{document}